# The Dynamics of Nestedness Predicts the Evolution of Industrial Ecosystems

Sebastián Bustos[1,2], Charles Gomez[3] & Ricardo Hausmann[1,2,4], César A. Hidalgo[1,5,6,†]

[1] Center for International Development, Harvard University
[2] Harvard Kennedy School, Harvard University
[3] Program on Organization Studies, Stanford University
[4] Santa Fe Institute.
[5] The MIT Media Lab, Massachusetts Institute of Technology
[6] Instituto de Sistemas Complejos de Valparaiso

† hidalgo@mit.edu

**Abstract:**

**In economic systems, the mix of products that countries make or export has been shown to be a strong leading indicator of economic growth. Hence, methods to characterize and predict the structure of the network connecting countries to the products that they export are relevant for understanding the dynamics of economic development. Here we study the presence and absence of industries at the global and national levels and show that these networks are significantly nested. This means that the less filled rows and columns of these networks' adjacency matrices tend to be subsets of the fuller rows and columns. Moreover, we show that nestedness remains relatively stable as the matrices become more filled over time and that this occurs because of a bias for industries that deviate from the networks' nestedness to disappear, and a bias for the missing industries that reduce nestedness to appear. This makes the appearance and disappearance of individual industries in each location predictable. We interpret the high level of nestedness observed in these networks in the context of the neutral model of development introduced by Hidalgo and Hausmann (2009). We show that, for the observed fills, the model can reproduce the high level of nestedness observed in these networks only when we assume a high level of heterogeneity in the distribution of capabilities available in countries and required by products. In the context of the neutral model, this implies that the high level of nestedness observed in these economic networks emerges as a combination of both, the complementarity of inputs and heterogeneity in the number of capabilities available in countries and required by products. The stability of nestedness in industrial ecosystems, and the predictability implied by it, demonstrates the importance of the study of network properties in the evolution of economic networks.**

**Introduction**

One of the best-documented findings of biogeography is that rare species inhabit predominantly diverse patches, while ubiquitous species tend to inhabit both, diverse and non-diverse locations[1-4]. In ecology, the term *nestedness* is used to refer to this feature, which has been observed numerous times in geographic patterns[1-4] and mutualistic networks[5-8]. In the case of mutualistic networks, nestedness implies that ecosystems are composed of a core set of interactions to which the rest of the community is attached[5]. The nestedness of interaction networks also implies that specialist species interact mostly with generalist species, and because generalist are less fluctuating[9], nestedness can help enhance the survival of rare species [10]. Nestedness has also been shown to enhance biodiversity[11] and overall ecosystem stability[12], and therefore, it is considered an important structural property of interaction networks in ecology.

Nestedness, however, is a general network measure that can be used to characterize non-biological ecosystems, such as global and national economies. In fact, the nestedness of economic systems has been described for interaction networks, connecting industries to other industries, such as the input-output matrices introduced half a century ago by Leontief[13], or the supply relationships in the New York Garment industry[14,15].

Here, we study the dynamics of economic geographic, instead. We look at the presence and absence of industries across a wide range of locations and show that (i) nestedness tends to remain stable; (ii) it can be used to predict the location of industrial appearances and disappearances; and (iii) can be accounted for by a simple model.

In recent years, the structure of industry-location networks has received a wide range of attention. A country's level of income is tightly connected to the mix of products that they export[16-18], as measured by the Economic Complexity Index or ECI[16,17]. The ECI is a structural measure of the network connecting countries to the products that they export that estimates the amount of productive knowledge embedded in a country[16] from information on who exports what. Countries that have an income that is lower than what would be expected from their ECI, such as China, India and Thailand, tend to grow faster than those that have an income that exceeds what would be expected from their current level of economic complexity, such as Greece and Portugal[16,17]. Hence, what countries export, as proxied by the ECI, is a strong leading indicator of economic growth.

The network connecting countries to the products that they export has been used to identify related varieties[19-21]. Here, products that tend to co-exported from the same location are connected with a strength that grows with the probability of co-export. Colocation networks, like the product space[20], have been used to show that the productive structure of countries and regions evolves as these diffuse towards products that are close, as measured by this network, to those that are already present in each location. The use of colocation data provides an alternative to more data intensive methods, such as networks connecting industries based on labor flows, labor similarities[22] or plant level data[23]. This is because labor and plant level data lacks standardized international coverage and therefore cannot be used to study global patterns and dynamics.

The evolution of a country's product mix, however, is highly path dependent[16,20]. Here, we look at the nestedness of the industry – location network and show that deviations from nestedness can help predict these path dependencies for both, industrial appearances and disappearances. These predictions add to our ability to explain the evolution of a country's product mix, and therefore, variations in cross-country levels of income. Moreover, we show that the high level of nestedness observed in the data, at the observed levels of fill of the matrix, can be reproduced using a simple model when we assume that the heterogeneity of capabilities available in a country, or required by a product, is large.

The paper is structured as follows. First, we study the nestedness of the industry - locations matrix and find it to be highly stable over time in spite of the fact that the density of the network increased by over 60 percent. We show this by using Almeida-Neto et al's NODF[24,25] (and Atmar and Patterson's Temperature metric[26,27] in the SM). We assess the stability of nestedness by comparing it with both, static and dynamic null models, showing that the observed level, and stability of the network's nestedness, is larger than what would be implied by these null models.

Next, we show that deviations from nestedness are associated, respectively, with increases and decreases in the probability that an industry will appear or disappear at a given location. Finally, to provide an explanation of the observed phenomena we generalize the model recently introduced by Hidalgo and Hausmann[16,28] that emphasizes the complementarity of inputs to show that this model can account for both, the high level of nestedness values, and their stability.

Together, these results illustrate the relevance of nestedness for the evolution of industrial ecosystems and shows that a simple model can account for the high level of nestedness observed in economic networks.

**Data & Methods**

The ideal data to study the patterns of economic geography would consist of plant level information, collected for all countries, with high spatiotemporal resolution, and following a disaggregates standardized classification covering all economic sectors. Unfortunately, such data is not available. Instead, we use yearly trade data connecting 114 countries to 772 different products. Here, products are classified according to the SITC-4 rev2 classification. We use data from 1985 to 2009 to approximate the evolution of the global patterns of production. We note that this consists of two different datasets, one that goes from 1985 to 2000[29], and another one that starts in 2001. The datasets do not match perfectly, and this boundary adds an additional source of variation to the data. Unfortunately, no consistent data source exists for the entire observation period. Going forward, we refer to this as the country-product network. We consider a country to be connected to a product if that country's exports per capita are larger than 25% of the world's exports per capita in that product for at least five consecutive years. These thresholds reduce the noise in the country – product data coming from re-exports and helps make sure that a country is connected to the products that it exports substantially and consistently. In the Supplementary Material we check for the robustness of our results by using a different definition of presences and absences based on Balassa's[30] Revealed Comparative Advantage (RCA), and find the results to be robust to this alternative definition of presences.

We note two important limitations of international trade data. First, it does not include products that are produced and consumed domestically. This is because it only considers a product once it has crossed an international border. Second, trade data is limited to goods, and therefore does not include any data on services. Despite these limitations, trade data is good for international comparisons because it is collected in a standardized classification that makes data for different countries comparable.

At the domestic level we use information on the tax residence of Chilean firms collected by Chile's *Servicio de Impuestos Internos* (SII), which is the equivalent of the United States Internal Revenue Service (IRS). Going forward, we refer to this dataset as the municipality-industry network. The municipality-industry network contains information on 100% of the firms that filed value-added and/or income taxes in Chile between 2005 and 2008. This data comprises firms from all economic sectors, whether they export or not, and whether they produce goods or services. The municipality-industry network consists of the universe of Chilean firms (nearly 900,000), which are classified into 700 different industries and assigned to each of Chile's 347 municipalities. Here we consider an industry to be present in a municipality if one or more firms, filing taxes under that industrial classification, declare that municipality as their tax residency.

Finally, we note that the Chilean tax data has the limitation that the tax residency of a firm can differ from the location of all of its operations. Going forward, we take the fact that our results hold in both, international trade and domestic tax data, as an indication that they are not driven by the limitations of these datasets and that they represent a natural characteristic of the economic networks underlying them. For more details on both datasets see the SM.

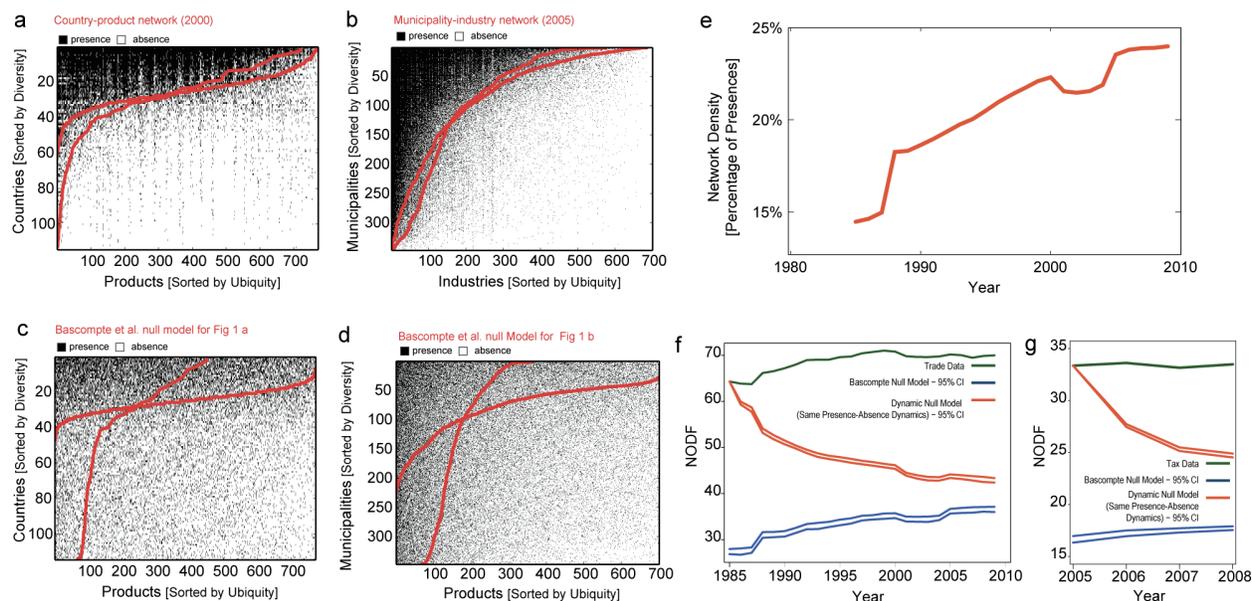

**Figure 1** The nestedness of international and domestic economies. **a** Country-product network for the year 2000. **b** Municipality-industry network for the year 2005. **c** Bascompte et al. null model for the matrix shown in **a**. **d** Bascompte et al. null model for the matrix presented in **b**. In **a-d** red lines indicate the diversity of a location and the ubiquity of an industry (see full text for details). **e** Evolution of the density, or fill, of the country-product network between 1985 and 2009. **f** Evolution of the NODF of the country-product network between 1985 and 2009 (green), its corresponding Bascompte et al. null model (blue, upper and lower lines indicate 95% conf. intervals), and that of a matrix that started identical to that for 1985, but that was evolved by considering an equal number of appearances and disappearances than in the original data (red, upper and lower lines indicate 95% conf. interval). **g** Same as **f** but for the municipality-industry network (see SM for results with Atmar and Patterson's temperature metric).

## Results

Figures 1 **a** and **b** show the matrices of the country-product and the municipality-industry networks (Respectively NODF=70.81 and NODF=83.35. We note that NODF=100 indicates perfect nestedness and NODF=0 indicates no nestedness)[31]. Here, the red lines indicate the diversity of each country and the ubiquity of each product -the number of locations where it is present- (see SM). These lines are used as a guide to indicate where presences would be expected to end if the nestedness of these networks were to be perfect. They can be thought of as a simplified extinction line[27]. Figure 1 **c** and **d** show their corresponding Bascompte et al. null models[5]. In the Bascompte et al. null model, the probability to find a presence in that same cell of the matrix is equal to the average of the probability of finding it in that row and column in the original matrix. The figures show that nestedness of the original networks is clearly larger than that of their respective null models, showing that industrial ecosystems are more nested than what would be expected for comparable networks (respective null model NODF of 35.0±0.6 and 46.5±0.3, errors are 99% conf. intervals calculated from 100 implementations of the null model).

Next, we study the temporal evolution of nestedness. In the case of the country-product network, where a larger time series is available (1985-2009), the percentage of presences almost doubled during the observation period (Figure 1 **e**), going from less than 15% to nearly 25%. In the case of the municipality-industry network, presences went up from 22.9% to 25.7% between 2005 and 2008. The nestedness of both, the country-product and the municipality-industry networks, however, remained relatively stable during this period as measured by NODF (green lines in Figure 1 **f-g** and SM).

We test the stability of these networks' nestedness by comparing it with two null models. The first one is an ensemble of null models[5] calculated for each respective year (blue lines in Figure 1 **f-g**). This shows that the nestedness of the empirical networks is always significantly higher than their randomized counterpart. Then, we show that a network subject to the same turnover dynamics would have lost its nestedness during the observation period. We do this by starting with the empirically observed network and simulate its evolution by sequentially adding and subtracting a number of links equal to the one gained or lost by the original network. We do this following the probability distributions defined by the Bascompte et al null model[5] to make sure that these additions and subtractions keep the degree sequence of the network close to the original one. Otherwise, the loss of nestedness could be a consequence of changes in the underlying distributions. This dynamic null model represent a strong control, since it preserves the exact density of the network and also its turnover dynamics, as the number of links that appear and disappear each year, in each country, and for each product is exactly that observed in the original data. The dynamic model, however, does not preserve nestedness, showing that its stability comes from the specific way in which links appear and disappear from the network, and not due to more trivial dynamics. In fact, when the appearance and disappearance of the links are chosen differently, the nestedness of the network quickly evaporates (red line in Figure 1 **f-g**). This allows us to conclude that the stability of nestedness observed in these networks is higher than what would be expected from a null model with the same general turnover dynamics.

Could the stability of nestedness be used to predict appearances and disappearances? In the literature, nestedness has been used to predict the biota available in ecological patches, albeit not in economic networks[2,32]. For the country-product network we define an appearance as an increase in exports per capita from less than 5% of the world average to more than 25%. To make sure that we are capturing structural changes and not mere short-term fluctuations, we ask that the increase in exports per capita of a country to go from less than 5%, for five consecutive years, to more than 25% sustained for at least 5 years. Hence, our final year of observation is 2005. Conversely, we count disappearances as a decrease in exports per capita of a country from 25% or more of the world's average to 5% or less (also sustained for at least 5 years). For the municipality-industry network we count appearances as changes from zero industries to one or more, and disappearances as changes from one or more industries to zero.

Figure 2 **a-d** visualizes the position in these networks' adjacency matrices of the industries that were observed to appear (green) and disappear (orange) in the intervening period. We predict these appearances and disappearances by fitting each observation in the industry-location network using a probit model that considers information on the diversity of the location and the ubiquity of the industry for the initial year (see SM). This represents a parameterization of nestedness and is similar to previous approaches that have used nestedness to make predictions[2,32]:

$$M_{c,p,t} = \alpha k_{c,t} + \beta k_{p,t} + \gamma(k_{c,t} \times k_{p,t}) + \epsilon_{c,p,t} \qquad (1)$$

Here $M_{c,p,t}$ is the industry-location network's adjacency matrix, $k_{c,t}$ is the diversity of location $c$ at time $t$ (defined as its degree centrality or $k_{c,t} = \sum_p M_{c,p,t}$), $k_{p,t}$ is the ubiquity of product $p$ at time $t$ (defined as its degree centrality or $k_{p,t} = \sum_c M_{c,p,t}$), and where we have also added an interaction term taking the product between diversity ($k_{c,t}$) and ubiquity ($k_{p,t}$). The error term is represented by $\epsilon_{c,p,t}$. We find that all coefficients are highly significant, meaning that a model that would only consider diversity or ubiquity, or both of them without an interaction term, would not be as accurate.

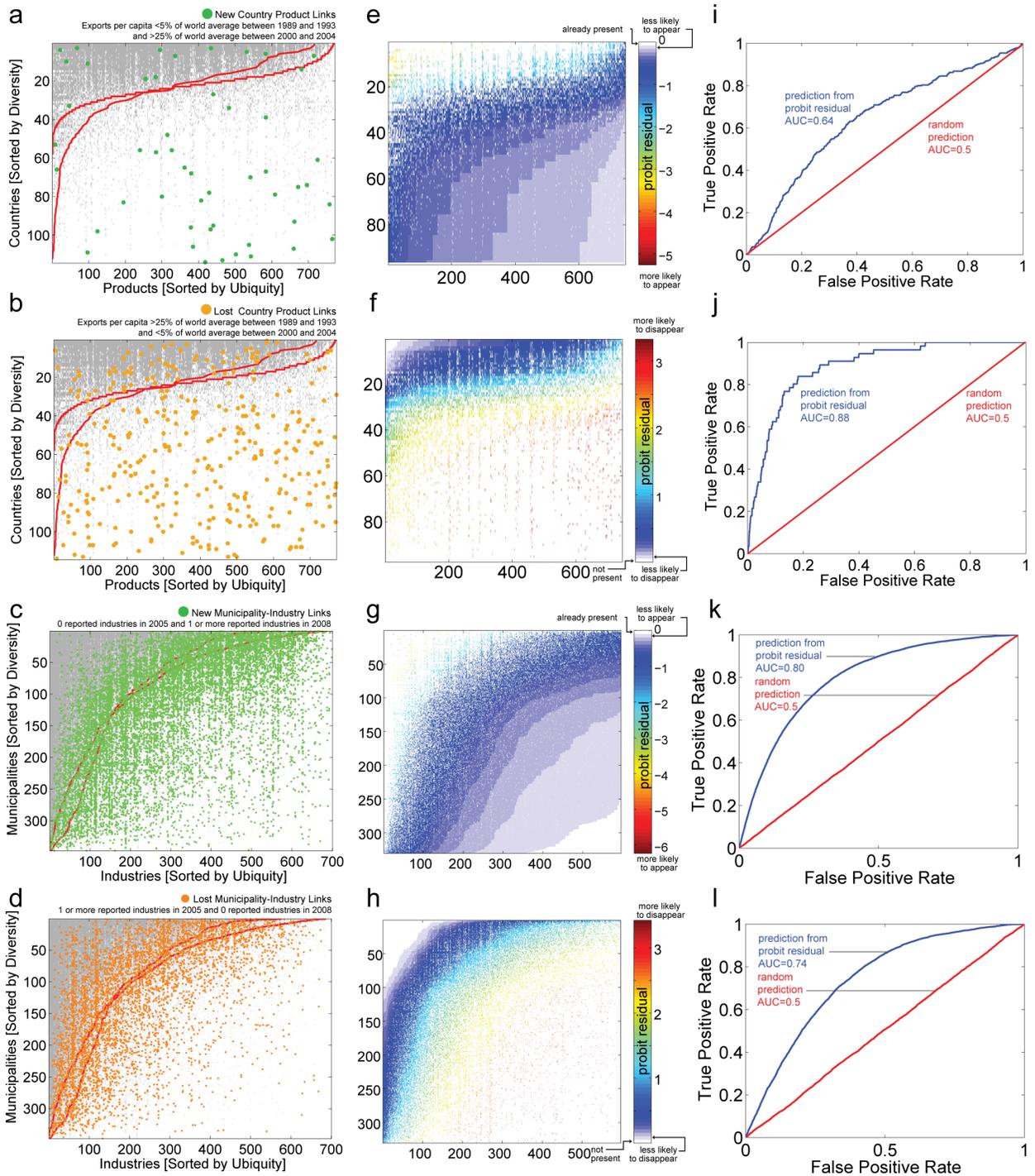

**Figure 2** Nestedness predicts appearing and disappearing industries. **a** The country-product network for the year 1993 is shown in grey. Green dots show the location of industries that were observed to appear between 1993 and 2000. **b** Same as **a**, but with the industries that disappeared in that period shown in Orange. **c** The municipality-industry network is shown in grey and green dots show the location of industries that were observed to appear between 2005 and 2008. **d** Same as **c**, but with the industries that disappeared in that period shown in Orange. **e-h** Deviance residuals of the regression presented in (1) applied to the presences-absences shown in **a-d**. **i-l** ROC curves summarizing the ability of the deviance residuals shown in **e-h**, to predict the appearances and dissapearences highlighted in **a-d**.

In general, we find that the probit regression fits presences and absences fairly accurately (average Efron's pseudo-$R^2$=0.53±0.02 for the country-product network and 0.54±0.01 for the municipality-industry network). Here, however, we use the deviance residuals of this regression to predict future appearances and disappearances. Negative residuals, represent unexpected absences[2] and are used to rank candidates for new appearances. Positive deviance residuals, on the other hand, represent unexpected presences[2] and are used to rank the likelihood that an industry will disappear in the future. (Figures 2 **e-h**).

But how accurate are these predictions? We quantify the accuracy of predictions by using the area under the Response Operator Characteristic curve or ROC curve[33,34]. An ROC curve plots the true positive rate of a prediction as a function of its false positive rate. The Area Under the Curve, or AUC, is commonly used to measure the accuracy of the prediction criterion[33,34]. A random prediction will find true positives and false positives at the same rate, and therefore will give an AUC of 0.5. A perfect prediction, on the other hand, will find all true positives before hitting any false positive and will be characterized by an AUC=1. Figures 2 **i-l** show the ROC curves obtained when the appearances and disappearances shown in Figures 2 **a-d** are predicted using the deviance residuals obtained from (1) for data on the initial year. In all cases, the ROC curves of these predictions (in blue), have an area that is significantly larger than the one expected for a random prediction (in red), showing that nestedness can help predict which links in these industry-location networks are more likely to appear or disappear.

Finally, we extend this analysis to all pairs of years. Figures 3 **a** and **b** show the number of events (appearances or disappearances) for each pair of years for the international trade data. As expected, there are fewer events for pairs of years that are close by in time. Also, we note that the number of appearances is larger than that of disappearances, a fact that is consistent with the observed increase in the density of the network. Figure 3 **c** shows the AUC value obtained for each pair of years, showing that for the country product network, disappearances (Fig 3 **b**) are predicted much more accurately than appearances.

The time series data available for Chile's municipality-industry network is much more limited. Hence, we show the average number of events (Figure 3 **d**), and the average AUC for networks separated by a given number of years (Figure 3 **e**). Here, we find that predictions of appearances and disappearance are both remarkably strong, and that there is no statistically significant difference in the predictability of both kinds of events.

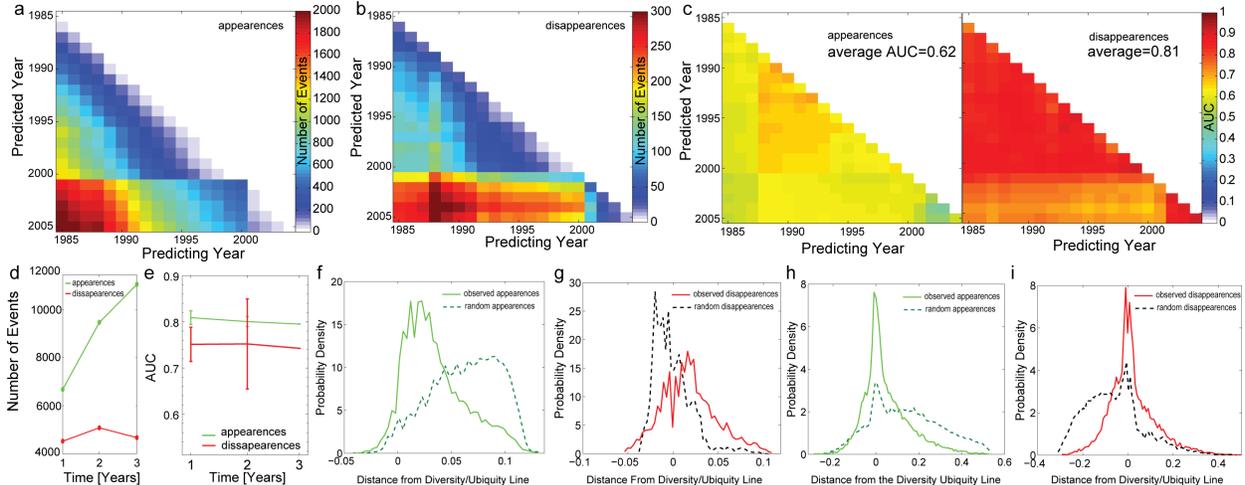

**Figure 3** Predicting appearances and disappearances using nestedness. **a** Number of appearances for every pair of years in the country-product network. **b** Number of disappearances for every pair of years for the country-product network. **c** Accuracy of the predictions for each pair of years measured using the Area Under the ROC Curve (AUC). **d** Average number of appearances and disappearances for the Chilean data (error bar smaller than symbol). **e** Average accuracy of the predictions for the municipality-industry network. Error bars indicate 99% confidence intervals. **f** Distribution for the distance to the diversity-ubiquity line obtained for the observed appearances and for an equal number of random appearances. **g** Same as **f** but for disappearances. **h** Same as **f**, but for the municipality-industry network. **i** Same as **h** but for disappearances.

To conclude this section, we look at the position in the network's adjacency matrix of appearances and disappearances. If the stability of nestedness is related to the location in this matrix of industrial appearances and disappearances, then appearances should be closer to the diversity-ubiquity line than random appearances. By the same token, disappearances should be farther away. For each event, we estimate its distance to the diversity and the ubiquity lines illustrated in figures 1 **a-d** and figures 2 **a-d** using

$$D = \text{Sign}((c',p')_i) \min\left(\left|\frac{\vec{I_p} - (c',p')_i}{N_c}\right|, \left|\frac{\vec{I_c} - (c',p')_i}{N_p}\right|\right). \quad (2)$$

Here $\vec{I_c}$ and $\vec{I_p}$ are respectively the lines of diversity and ubiquity (i.e. the red lines in Figure 1 a-d), $(c',p')_i$ is the position in the adjacency matrix of the $i^{th}$ event, and $N_c$ and $N_p$ are respectively the number of locations and industries in the network. We use $N_c$ and $N_p$ to normalize the maximum possible vertical and horizontal distances to 1 and thus make sure that the measure is less sensitive to the rectangularity of the different matrices. The || operator represents the Euclidean distance and $\text{Sign}((c',p')_i) = 1$ if the position of the event is outside of the nested area defined by both $\vec{I_c}$ and $\vec{I_p}$ and -1 otherwise (see SM). As a benchmark comparison we consider an equal number of appearances and disappearances, but draw these from a random set of eligible positions in the adjacency matrix.

Figures 3 **f-i** compare the distributions of distances ($D$) with those associated with an equal number of random appearances or disappearances. We find that appearances tend to lie significantly closer to the diversity/ubiquity lines than what would be expected for an equal number of random events (ANOVA F=59,935, p-value=0 for the country-product network and ANOVA=10895 p-value=0 for the municipality-industry network). In the case of disappearances, the opposite holds true. The observed appearances tend to be mostly located outside of the nested area defined by the diversity/ubiquity lines. By chance, however, disappearances would come mostly from the highly populated area inside the diversity/ubiquity lines. Once again, differences between observations and null model expectations are highly significant for both networks (ANOVA F=6246 p-value=0 for the country-product network and ANOVA F=6463 p-value=0 for the municipality-industry network).

Finally, we show that a modified version of the neutral development model introduced in [17], and solved analytically in [28], can be used to explain both, the observed level of nestedness and its stability. This neutral development model consists of three simple assumptions.

(i) Products require a set of non-tradable inputs, or capabilities, to be produced.
(ii) Locations are characterized by a set of capabilities.
(iii) Locations can only produce the products for which they have all the required capabilities.

The model is formalized by introducing three mathematical objects: two matrices and one operator. $P_{pa}$ is a matrix that is 1 if product $p$ requires capability $a$ and 0 otherwise. $C_{ca}$ is a matrix that is 1 if location $c$ has capability $a$, and zero otherwise. Finally (iii) provides a way of mapping $C_{ca}$ and $P_{pa}$ into $M_{cp}$, since it implies that $M_{cp} = 1$ if the set of capabilities required by a product is a subset of the capabilities available in a location. Mathematically (iii) can be expressed as the following operator:

$$M_{cp} = 1 \; if \; \sum_a P_{pa} = \sum_a C_{ca}P_{pa} \; and \; M_{cp} = 0 \; otherwise. \tag{3}$$

More details about the model can be found in [28].

To compare the model to the data we need to assume the form of $C_{ca}$ and $P_{pa}$. In [28] the model was solved analytically by assuming that both, $C_{ca}$ and $P_{pa}$ were random matrices. This means that each location has a capability with probability $r$ and that products require a capability with a probability $q$. From this we can trivially deduce that the number of capabilities available in a random country, or required by a random product, follows a binomial distribution. Because of this, we call this implementation of the neutral model: the binomial model. The third and final parameter that needs to be specified is the number of capabilities required by a product ($N_a$). This is because the number of locations $N_c$, and the number of products $N_p$, is fixed to match the number of locations and products observed in the data.

Effectively, the binomial model has two free parameters. This is because it is always possible to determine $r$, $q$ or $N_a$ once the fill of the $M_{cp}$ matrix is known. The binomial model has been shown to reproduce the distribution of diversities, ubiquities, co-exports, and the relationship between diversity and ubiquity of the country-product network using $N_a$=80, $r$=0.87 and $q$=0.18. In addition to the binomial model we consider an alternative form that has the same number of parameters. We call this the uniform model, since in this case the number of capabilities that a country has is distributed uniformly between 0 and $R$ and the number of capabilities that a product requires is distributed uniformly between 0 and $Q$. Hence, in this model country $c$ has a capability $a$ with probability equal to $r_c$=min(1,R×c/$N_c$).

We take the minimum to ensure $r_c$ is upper bounded by 1. In the uniform model, allowing values of $R$ larger than one allows having a small number of fully diversified countries.

Figures 4 a and 4 b illustrate the binomial model and the uniform model, respectively. For both models, we show their respective $C_{ca}$ and $P_{pa}$ matrices together with their resulting country-product network $M_{cp}$. We find that in both cases the resulting $M_{cp}$ matrices are significantly more nested than the null model, yet the nestedness emerging from the uniform model is considerably larger, resembling closely the values observed for the country-product network. This comes from the fact that countries with a diverse capability endowment are likely to make a wide range of products, whereas countries with few capabilities will only be able to make those products that require few capabilities. This last observation is implied by assumption (iii), and is therefore true for both, the binomial and the uniform model. Yet, the large degree of heterogeneity among countries and products present in the uniform model enhances the nestedness implied by the complementarity assumption.

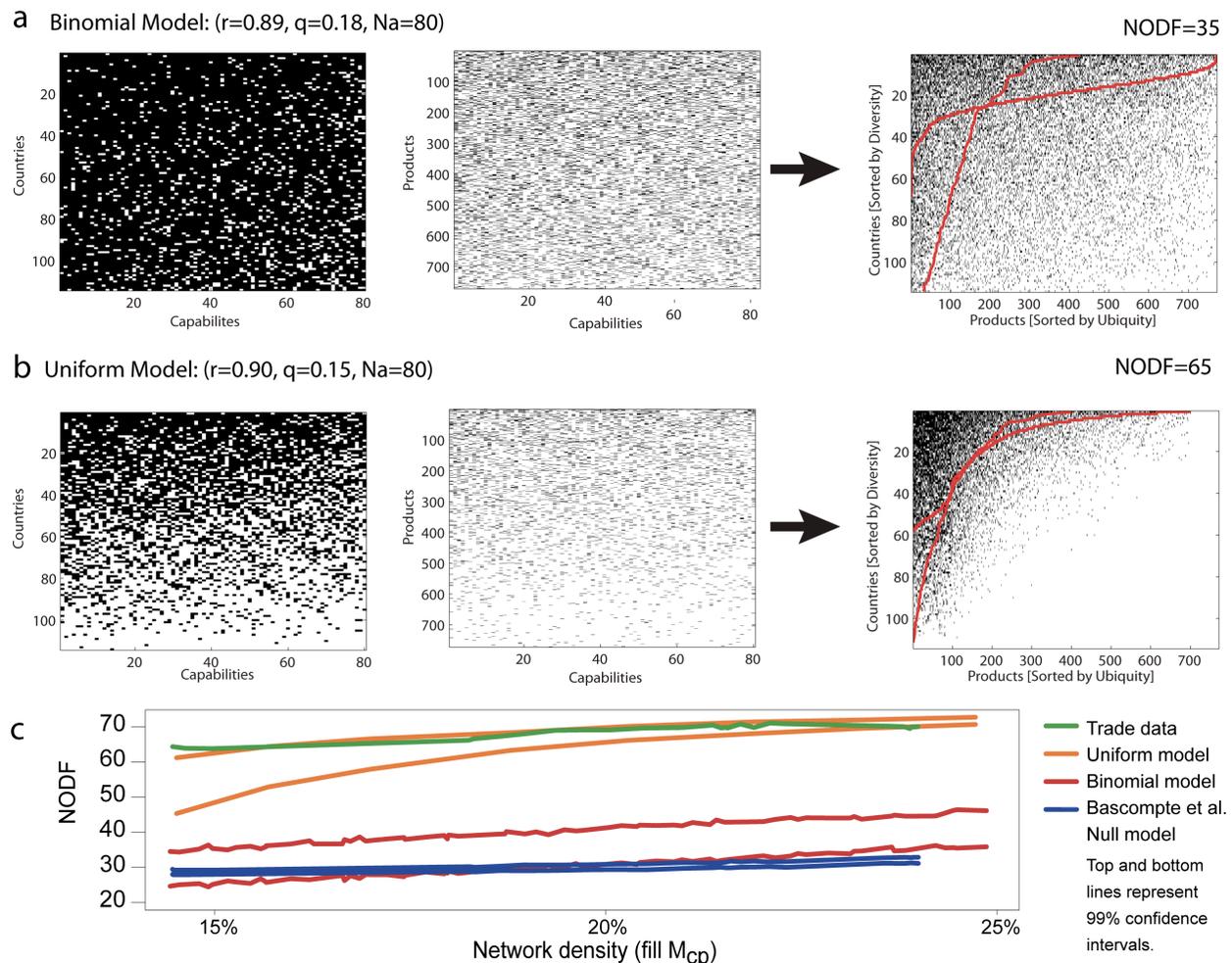

**Figure 4** Modeling nestedness. **a** Illustration of the binomial model. From left to right $C_{ca}$, $P_{pa}$ and the resulting $M_{cp}$. **b** Illustration of the uniform model. From left to right $C_{ca}$, $P_{pa}$ and the resulting $M_{cp}$. **c** NODF as a function of matrix fill for the country-product network (green), the uniform model (orange), the binomial model (red), and the Bascompte et al null model (blue).

Figure 4 c compares the nestedness of the country-product network with the one found for the neutral models and null model. Here we plot nestedness as a function of the fill of the network since this is a good proxy for time and the neutral models and null model do not have an explicit time dimension. We implement this comparison by generating an initial $P_{pa}$ matrix that is kept constant during the procedure. In the binomial model we choose $q$=0.18, and for the uniform model we take $Q$=0.21. We interpret this as an assumption that productive technologies change slowly during the time frames considered, and therefore, the increases in diversification observed in the empirical network comes from locations developing capabilities and catching up to produce the products that more diversified locations were already making. To create $M_{cp}$, we generate 100 $C_{ca}$ matrices for 200 different values of $r$ and $R$. For the binomial model we consider values of $r$ between 0.9 and 0.95, while for the uniform model we consider values of $R$ between 0.9 and 1.07. In both cases we set the total number of capabilities in the system to $N_a$=80. These values are chosen to ensure that the fills of the modeled $M_{cp}$ matrices are close to the ones observed in the original data. The analysis shows that the nestedness of the $M_{cp}$ matrices implied by the neutral model matches the ones observed in the economic networks only for the uniform model. In the context of assumptions (i)-(iii), we interpret this result as evidence that heterogeneity in the distribution of capabilities available in a country, or required by a product, are needed to generate the high levels of nestedness observed in these economic networks.

**Discussion**

In this paper we showed that industry-location networks are nested, just like industry-industry networks[13-15], or their biological counterparts[1-4,26,27]. Using time series data for both, international and domestic economies, we showed that the nestedness of these networks tends to remain stable over time and that this empirical regularity can be used to predict the pattern of industrial appearances and disappearances. Moreover, we showed that, for the empirically observed fills, we can account for the high level of nestedness observed in the world using a simple model, but only if we assume a relatively large degree of heterogeneity in the number of capabilities present in a country or required by a product.

The strong link between biological and industrial ecosystems opens a variety of questions. First, is the geographical nestedness described in this paper a consequence of industry-industry nestedness, or are these independent phenomena? Second, are the mechanisms generating nestedness at the global level the same that generate nestedness at the national level?

In this paper we showed that the geographical nestedness of industries holds at both, the global and at the national scale. This is certainly not the case for biological ecosystems, since the biota of the artic is not a subset of that of the rain forest. The fact that the nestedness of industrial ecosystems holds at scales as large as that of the world economy suggests that the coupling between international economies is strong. For instance, most products could potentially be produced anywhere in the world, if the right industrial environment would be available. Yet this is probably not true for species, which are more dependent on climate and geography. In the context of the proposed model, this is expressed in the fact that all countries share access to the same $P_{pa}$ matrix, so they are able to create a product if they acquire the requisite capabilities.

The predictability implied by nestedness, on the other hand, has important implications in a world where income per capita is connected to the mix of products that a country makes[17,18]. Ultimately, the dynamics that underpin nestedness imply certain dynamics in the way countries diversify and constrain the speed at which the income gaps between rich and poor countries can decline.

More research will certainly need to be done on both, the causes of the structures and the time patterns that were uncovered in this paper. This will require strengthening the bridge between the natural and social sciences because, if there is something that the nestedness of economies show, is that humans tend to generate patterns in social systems that strongly mimic those found in nature[35,36].

**ACKNOWLEDGEMENTS:**
We acknowledge the support of the Santo Domingo Foundation, Standard Bank, the ABC Career Development Chair at the MIT Media Lab, and the MIT Media Lab Consortia. We also thank Jordi Bascompte, Frank Neffke, Javier Galeano, Pablo Marquet, Daniel Stock & Muhammed Yildirim for their comments.

**References:**

1. Hulten, E. *Outline of the history of Artic and Boreal biota during the Quaternary Period*, Lund University, (1937).
2. Ulrich, W., Almeida, M. & Gotelli, N. J. A consumer's guide to nestedness analysis. *Oikos* **118**, 3-17, doi:10.1111/j.1600-0706.2008.17053.x (2009).
3. Darlington, P. J. *Zoogeography: the geographical distribution of animals.*, (wiley, 1957).
4. Daubenmire, R. Floristic plant geography of eastern Washington and northern Idaho. *Journal of Biogeography* **2** (1975).
5. Bascompte, J., Jordano, P., Melian, C. J. & Olesen, J. M. The nested assembly of plant-animal mutualistic networks. *Proceedings of the National Academy of Sciences of the United States of America* **100**, 9383-9387, doi:10.1073/pnas.1633576100 (2003).
6. Dupont, Y. L., Hansen, D. M. & Olesen, J. M. Structure of a plant-flower-visitor network in the high-altitude sub-alpine desert of Tenerife, Canary Islands. *Ecography* **26**, 301-310, doi:10.1034/j.1600-0587.2003.03443.x (2003).
7. Ollerton, J., Johnson, S. D., Cranmer, L. & Kellie, S. The pollination ecology of an assemblage of grassland asclepiads in South Africa. *Annals of Botany* **92**, 807-834, doi:10.1093/aob/mcg206 (2003).
8. Gilarranz, L. J., Pastor, J. M. & Galeano, J. The architecture of weighted mutualistic networks. *Oikos*, 001–009, doi:doi: 10.1111/j.1600-0706.2011.19592.x (2011).
9. Turchin, P. & Hanski, I. An empirically based model for latitudinal gradient in vole population dynamics. *American Naturalist* **149**, 842-874, doi:10.1086/286027 (1997).
10. Jordano, P. PATTERNS OF MUTUALISTIC INTERACTIONS IN POLLINATION AND SEED DISPERSAL - CONNECTANCE, DEPENDENCE ASYMMETRIES, AND COEVOLUTION. *American Naturalist* **129**, 657-677, doi:10.1086/284665 (1987).
11. Bastolla, U. *et al.* The architecture of mutualistic networks minimizes competition and increases biodiversity. *Nature* **458**, 1018-U1091, doi:10.1038/nature07950 (2009).
12. Bascompte, J., Jordano, P. & Olesen, J. M. Asymmetric coevolutionary networks facilitate biodiversity maintenance. *Science* **312**, 431-433, doi:10.1126/science.1123412 (2006).
13. Leontief, W. W. The Structure of the U.S. Economy. *Scientific American* **212**, 25-35 (1965).
14. Saavedra, S., Reed-Tsochas, F. & Uzzi, B. A simple model of bipartite cooperation for ecological and organizational networks. *Nature* **457**, 463-466, doi:10.1038/nature07532 (2009).
15. Saavedra, S., Stouffer, D. B., Uzzi, B. & Bascompte, J. Strong contributors to network persistence are the most vulnerable to extinction. *Nature* **478**, 233-U116, doi:10.1038/nature10433 (2011).
16. Hausmann, R., Hidalgo, C. A. & al., e. *The Atlas of Economic Complexity*. (Puritan Press, Cambridge MA., 2011).
17. Hidalgo, C. A. & Hausmann, R. The building blocks of economic complexity. *Proceedings of the National Academy of Sciences of the United States of America* **106**, 10570-10575, doi:10.1073/pnas.0900943106 (2009).
18. Hausmann, R., Hwang, J. & Rodrik, D. What you export matters. *Journal of Economic Growth* **12**, 1-25, doi:10.1007/s10887-006-9009-4 (2007).

19  Teece, D. J., Rumelt, R., Dosi, G. & Winter, S. UNDERSTANDING CORPORATE COHERENCE - THEORY AND EVIDENCE. *Journal of Economic Behavior & Organization* **23**, 1-30 (1994).
20  Hidalgo, C., Klinger, B., Barabasi, A. & Hausmann, R. The product space conditions the development of nations. *Science* **317**, 482-487 (2007).
21  Bryce, D. J. & Winter, S. G. A General Interindustry Relatedness Index. *Management Science* **55**, 1570-1585, doi:10.1287/mnsc.1090.1040 (2009).
22  Neffke, F. & Svensson Henning, M. *Revealed relatedness: Mapping industry space* (Working Paper Series 08.19, Papers in Evolutionary Economic Geography, Utrecht University,Utrecht, the Netherlands., 2008).
23  Neffke, F., Henning, M. & Boschma, R. How Do Regions Diversify over Time? Industry Relatedness and the Development of New Growth Paths in Regions. *Economic Geography* **87**, 237-265, doi:10.1111/j.1944-8287.2011.01121.x (2011).
24  Almeida, M., Guimaraes, P. R. & Lewinsohn, T. M. On nestedness analyses: rethinking matrix temperature and anti-nestedness. *Oikos* **116**, 716-722, doi:10.1111/j.2007.0030-1299.15803.x (2007).
25  Almeida-Neto, M., Guimaraes, P., Guimaraes, P. R., Loyola, R. D. & Ulrich, W. A consistent metric for nestedness analysis in ecological systems: reconciling concept and measurement. *Oikos* **117**, 1227-1239, doi:10.1111/j.2008.0030-1299.16644.x (2008).
26  Atmar, W. & Patterson, B. D. THE MEASURE OF ORDER AND DISORDER IN THE DISTRIBUTION OF SPECIES IN FRAGMENTED HABITAT. *Oecologia* **96**, 373-382, doi:10.1007/bf00317508 (1993).
27  Patterson, B. D. & Atmar, W. NESTED SUBSETS AND THE STRUCTURE OF INSULAR MAMMALIAN FAUNAS AND ARCHIPELAGOES. *Biological Journal of the Linnean Society* **28**, 65-82, doi:10.1111/j.1095-8312.1986.tb01749.x (1986).
28  Hausmann, R. & Hidalgo, C. A. The network structure of economic output. *Journal of Economic Growth* **16**, 309-342, doi:10.1007/s10887-011-9071-4 (2011).
29  Feenstra, R. C., Lipsey, R. E., Deng, H., Ma, A. C. & Mo, H. World Trade Flows:1962-2000. *NBER Working Paper No. W11040* (2005).
30  Balassa, B. COMPARATIVE ADVANTAGE IN MANUFACTURED GOODS - A REAPPRAISAL. *Review of Economics and Statistics* **68**, 315-319 (1986).
31  Guimaraes, P. R. & Guimaraes, P. Improving the analyses of nestedness for large sets of matrices. *Environmental Modelling & Software* **21**, 1512-1513, doi:10.1016/j.envsoft.2006.04.002 (2006).
32  Maron, M., Mac Nally, R., Watson, D. M. & Lill, A. Can the biotic nestedness matrix be used predictively? *Oikos* **106**, 433-444, doi:10.1111/j.0030-1299.2004.13199.x (2004).
33  Bradley, A. P. The use of the area under the roc curve in the evaluation of machine learning algorithms. *Pattern Recognition* **30**, 1145-1159, doi:10.1016/s0031-3203(96)00142-2 (1997).
34  Zweig, M. H. & Campbell, G. RECEIVER-OPERATING CHARACTERISTIC (ROC) PLOTS - A FUNDAMENTAL EVALUATION TOOL IN CLINICAL MEDICINE. *Clinical Chemistry* **39**, 561-577 (1993).
35  Bettencourt, L. & West, G. A unified theory of urban living. *Nature* **467**, 912-913 (2010).

36   West, G. B., Brown, J. H. & Enquist, B. J. A general model for the origin of allometric scaling laws in biology. *Science* **276**, 122-126 (1997).

**SUPPLEMENTARY MATERIAL FOR:**

# The Dynamics of Nestedness Predicts the Evolution of Industrial Ecosystems

**Sebastián Bustos**[1,2]**, Charles Gomez**[3] **& Ricardo Hausmann**[1,2,4]**,César A. Hidalgo**[1,5,6,†]

[1] Center for International Development, Harvard University
[2] Harvard Kennedy School, Harvard University
[3] Program on Organization Studies, Stanford University
[4] Santa Fe Institute.
[5] The MIT Media Lab, Massachusetts Institute of Technology
[6] Instituto de Sistemas Complejos de Valparaiso
[†] hidalgo@mit.edu

## TABLE OF CONTENTS

## DATA DETAILS:

### International Trade Data:

The international data set is a merge of two data sources: The Feenstra et al. (2005) data set, which has data for the years prior to 2000, and the UN Comtrade database (comtrade.un.org), which we used for the period going from 2001 to 2009. Both dataset follow the product classification established by the Standard International Trade Classification (SITC) revision 2[*]. In the UN Comtrade dataset we associated countries to products according to what was reported as exports to the WLD category (World). For the products in which no exports to WLD (World) was found, exports were reconstructed using the reports from importing countries, when available, and by aggregating the reported bilateral exports of the exporting country as a last resource. We prioritize imports over exports because imports tend to be more tightly controlled than exports.

While the Feenstra et. al (2005) data set contains trade starting 1962, we chose 1985 as our starting year because there are several reclassifications of the data that affect their reliability for previous years (see SM2 Data Continuity). Since presences are averages over 5 years, the first year that is included in our dataset is 1981 (in the counting of presences for 1985).

We find, however, that international trade data is characterized by a nested matrix even for the years that we do not include in this paper. Figure SM1 shows the Temperature and NODF calculated for all years. Our choice to restrict the number of years in the dataset was performed to reduce the number of false appearances and disappearances that could be introduced by reclassifications of the SITC categories.

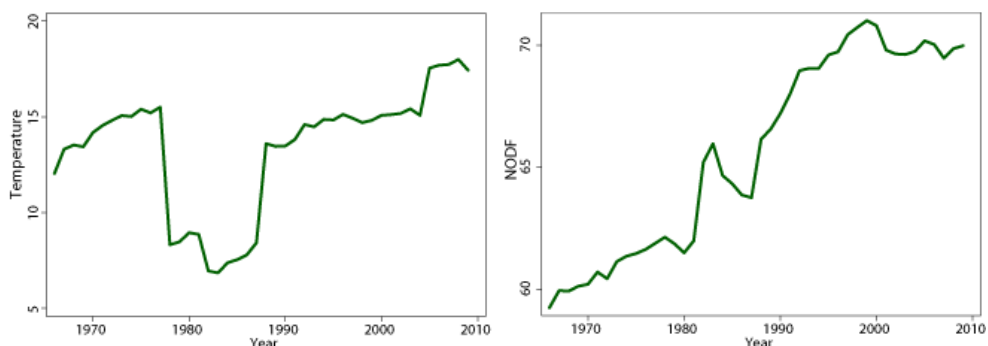

**Figure SM1** The nestedness of international trade data from 1966 to 2009. Note the range of the y-axis.

Finally, we restrict the sample of countries for all those that have a population of at least 1.25 million in the year 2000. We also remove countries from the Former Soviet Union (FSU), because these countries lack data for the 1980's, and have noisy data for the early 1990's. For Germany, we use data on West Germany for the years prior to 1992.

The final dataset consists of 114 countries and 775 products, classified according to the SITC4 rev2 classification (http://reportweb.usitc.gov/commodities/naicsicsitc.html).

---

[*] For more information visit http://unstats.un.org/unsd/cr/registry/regcst.asp?Cl=8&Lg=1

The datasets includes only tradable products, from raw materials and agriculture, to manufactures and chemicals.

**Domestic Tax Data:**

The domestic data for Chile consists of a matrix indicating the number of firms from a given industry in each municipality. The data has records for the year 2005, 2006, 2007 and 2008 and is based on the fiscal residence of each firm (it is hence a firm, and not a establishment level dataset). The number of firms reported for each year is shown in table 1.

| Year | Number of Firms |
|------|-----------------|
| 2005 | 862,405 |
| 2006 | 876,948 |
| 2007 | 891,383 |
| 2008 | 899,156 |

Table 1: Number of Firms in the Chilean Tax Data

These data contains information on the universe of Chilean firms and includes firms from all economic sectors, from raw materials and manufacturing, to restaurant, retail and banking services. The data contains information for 347 municipalities and 700 industries classified according to the Código the Actividad Económica (CAE) (http://www.sii.cl/catastro/codigos.htm).

**PRESENCE-ABSENCE MATRIX DEFINITION:**

For the international trade data set, we define the presences of an industry in a country if that country has exports per capita that are at least 25% of the world average for 5 consecutive years. Formally, we do this following:

$$M_{cp} = 1 \text{ if } \frac{EXP_{cp}/P_c}{\sum_c EXP_{cp}/\sum_c P_c} > 0.25 \text{ and } M_{cp} = 0 \text{ Otherwise}$$

Where $M_{cp}$ is the presence-absence matrix, $EXP_{cp}$ are the exports of product $p$ by country $c$, and $P_c$ is the population of country $c$. For the domestic tax data, we define as a presence a municipality that has one or more firms filing taxes under that industrial classification. We use a single year in this case.

**NESTEDNESS METRICS: TEMPERATURE AND NODF**

We calculate the nestedness of the exports per capita absence-presence matrices using both, Atmar and Patterson's temperature metric and Almeida-Neto et al.'s NODF metric. Preparation of these matrices for both analyses is similar. For the temperature metric, the rows and columns of a matrix are sorted and rank-ordered to yield a nested matrix with the absolute minimum temperature possible for this matrix. For the NODF metric, the rows and the columns of a matrix are swapped and rank-ordered by the sum of the presences in each of these rows and columns, respectively. The transformed matrices are then ready to be processed by the following algorithms. For a more detailed explanation, please reference the respective works of Atmar and Patterson (1993) and Almeida-Neto et al. (2008). Also the review by Ulrich, Almeida and Gotelli (2009) is a good place to learn about both of these metrics.

### ATMAR AND PATTERSON'S TEMPERATURE MEASURE

Atmar and Patterson's temperature metric calculates the number and the degree of unexpected presences and absences in an ordered adjacency matrix. Unexpected presences and absences are calculated with respect to an *extinction line* that separates the adjacency matrix into two areas: The top-left triangle, which we will call Section 1, where only presences are expected to appear, and the bottom right triangle, which we call Section 2, where only absences are expected (Figure SM2). In a perfectly nested matrix an ideal extinction line is a skew diagonal bisecting the matrix, where all of the presences are to one side of the line and all of the absences are to the other side.

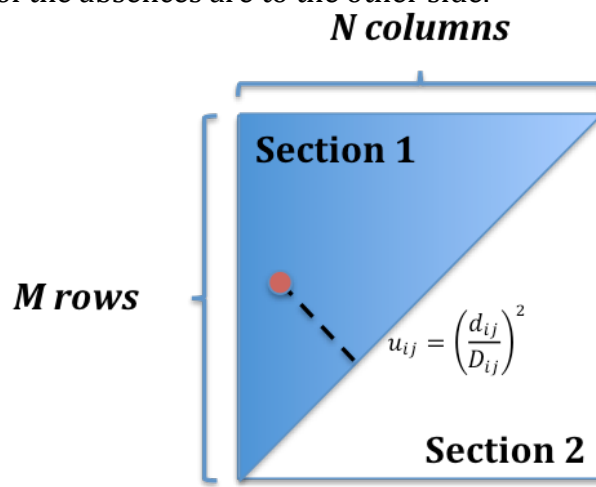

**Figure SM2:** A Perfectly Nested Matrix with M Rows and N columns

Presences in Section 1 that are closer to the extinction line are considered more likely to face extinction. A presence in Section 2, on the other hand, is considered an unexpected presence. Thus, the distance from the extinction line captures the degree of unexpectedness of presence. Conversely, absences in Section 1 are considered unexpected absences.

A perfectly nested matrix is characterized by a temperature of zero degrees. Alternatively, a fully disordered matrix is characterized by a temperature of 100 degrees.

The degree of "unexpectedness" for any presence or absence is the squared ratio of its distance, $d_{ij}$, to the ideal extinction line, $D_{ij}$. This local unexpectedness is expressed as:

$$u_{ij} = \left(\frac{d_{ij}}{D_{ij}}\right)^2$$

The degree of unexpectedness for the matrix, $U$, is the sum of each of these local unexpectedness values. This sum is normalized by the number of rows ($m$) and columns ($n$), to ensure the measure is unaffected by the size or the shape of the adjacency matrix:

$$U = \frac{1}{mn} \sum_{ij} u_{ij}$$

The total unexpectedness is transformed to a temperature scale using a normalization factor. The temperature scale goes from 0 degrees, corresponding to a perfectly ordered matrix, to 100 degrees, indicating a matrix full with unexpected values:

$$T = \frac{100}{U_{max}} U$$

where $U_{max}$ is 0.04145.

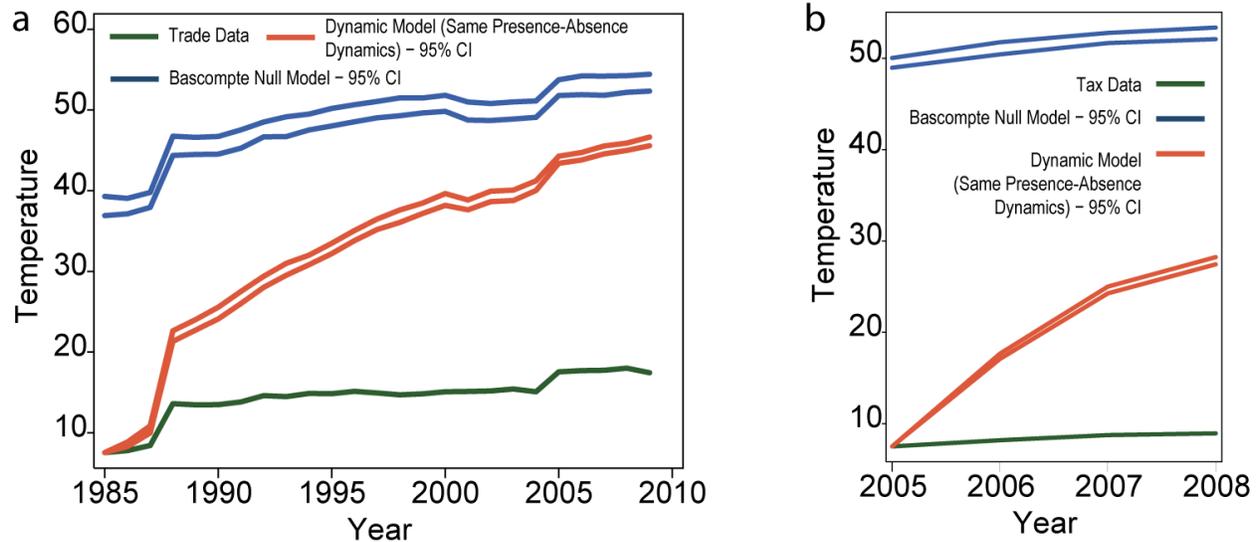

Figure **SM3**: The results shown in Figure 1 f and g of the main text but using temperature.

### Almeida-Neto et al.'s NODF Measure

The Nested Overlap and Decreasing Fill, or NODF metric, measures the degree of overlap between an adjacency matrix's rows and columns. NODF is determined by comparing all row-row and all column-column pairs. A row-row pair $ij$ is any row $i$ paired with each row above it, row $j$, in an ordered matrix. Similarly, a column-column pair $ij$ is any column $i$ paired with each column behind it, column $j$, in an ordered matrix. This is first achieved by calculating the Paired Overlap, $PO_{ij}$, for each row-row and each column-column pair. $PO_{ij}$ is calculated as the percentage of presences in row or column $i$ that are also present in row or column $j$:

$$PO_{ij} = \frac{\sum O_{ij}}{\sum MT_j}$$

where $MT_j$ is marginal total, the sum of presences in row or in column $j$, and $O_{ij}$ is the number of presences overlapping between the row-row or the column-column pair.

Figure **SM4**: An Ordered Matrix[†]

For example, consider rows r1 and r2 from Figure SM4:

Figure **SM4a**: A Sample Row-Row Pair from the Matrix in Figure SM3

In figure SM4a, r2 – the less populated row with three presences – overlaps with two presences in r1 – the more populated row. The $PO_{ij}$ for the r1-r2 pairing is thus two presences divided by three presences, or $PO_{12}$ = 66.67%. Similarly for columns, consider figure **SM4b**:

Figure **SM4b**: A Sample Column-Column Pair from the Matrix in Figure **SM4**

Column c4 – the less populated column with only two presences – shares only shared presence with column c1 – the more populated column. Thus, the $PO_{ij}$ for the c1-c4 pairing is $PO_{14}$ = 50%.

With the paired overlap, we can now calculate both the decreased fill, $DF_{ij}$, for every row-row and column-column pair. The $DF_{ij}$ takes one of two values depending on the marginal total, or $MT$, of the rows or the columns in the pair. Thus, in an ordered adjacency matrix, if the marginal total of row $i$, $MT_i$, is less than the marginal total of row $j$, $MT_j$, then $DF_{ij}$ takes on the value of 100. Otherwise, if $MT_i$ is greater than or equal to the $MT_j$, then $DF_{ij}$ takes on the value of 0.

---

[†] Figures SM3, SM3a, SM3b, and SM4 are taken directly from: *Almeida-Neto, M., Guimaraes, P., Guimaraes, P. R., Loyola, R. D. & Ulrich, W. A consistent metric for nestedness analysis in ecological systems: reconciling concept and measurement. Oikos* **117**, *1227-1239, doi:10.1111/j.2008.0030-1299.16644.x (2008).*

$$\begin{cases} DF_{ij} = 100, & MT_i < MT_j \\ DF_{ij} = 0, & MT_i \geq MT_j \end{cases}$$

The penultimate variable is the paired nestedness, $N_{ij}$, for every row-row and every column-column pair. Similar to $DF_{ij}$, $N_{ij}$ can take on only one of two values based on the $DF_{ij}$ and the $PO_{ij}$ of its row-row or its column-column pair. Thus, if $DF_{ij}$ = 100, then $N_{ij}$ = $PO_{ij}$; otherwise, $N_{ij}$ = 0.

$$\begin{cases} N_{ij} = PO_{ij}, & DF_{ij} = 100 \\ N_{ij} = 0, & Otherwise \end{cases}$$

The $N_{ij}$ is calculated for every row-row and column-column pair in the matrix. The NODF score is the average of all $N_{ij}$ values:

$$NODF = \frac{\sum N_{ij}}{\frac{n(n-1)}{2} + \frac{m(m-1)}{2}}$$

where $\frac{n(n-1)}{2}$ and $\frac{m(m-1)}{2}$ are the the total number of possible row-row and column-columns pairs in the matrix. Figure **SM4c** illustrates the entire NODF calculation for the matrix in figure 2.

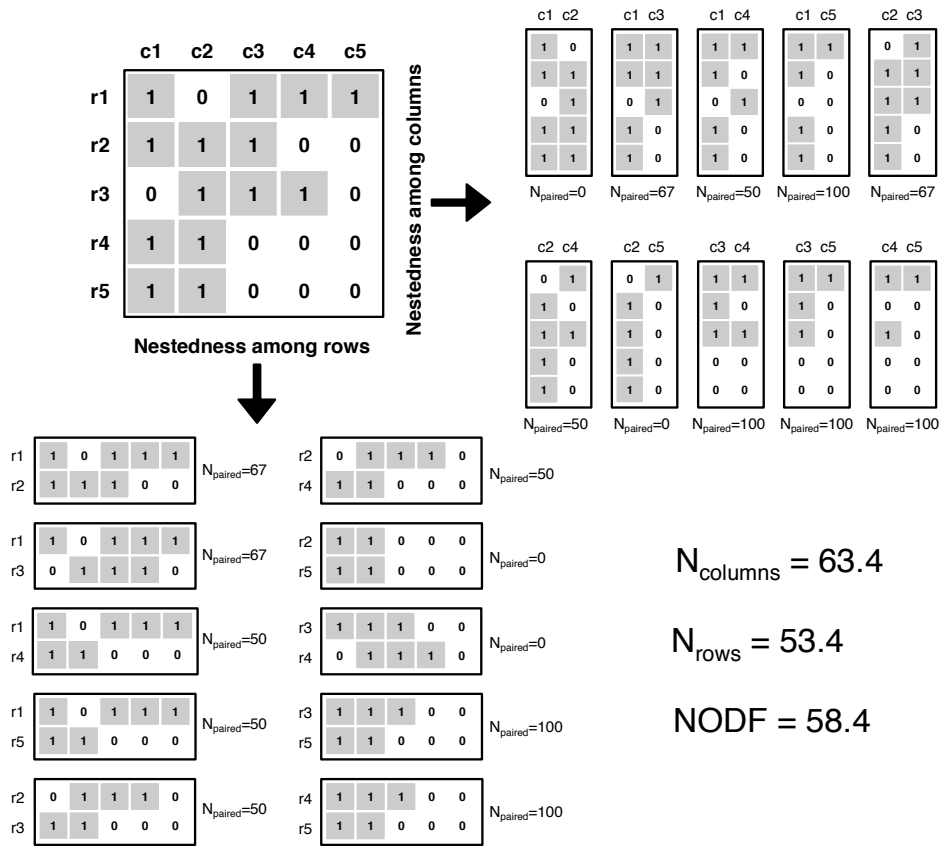

Figure **SM4c**: The Entire NODF Calculation for the Matrix in Figure 2

# NULL MODELS

## Static Null Model (Bascompte et al.)

Bacompte et al. (2003) introduced a null model to show whether the nested order of the data is statistically meaningful. For this, they introduced a null model ($M_{cp}*$) in which the probability to find a presence in that same cell of the matrix is equal to the average of the probability of finding it in that row and column in the original matrix ($M_{cp}$).

$$P(M_{cp}^* = 1) = \frac{1}{2}\left(\frac{1}{N_p}\sum_p M_{cp} + \frac{1}{N_c}\sum_c M_{cp}\right)$$

Using this model we performed 100 random realizations of the matrix for each year. Then we calculated the Temperature and NODF of each realization of the resulting null matrices to obtain a distribution of possible outcomes. Figures 1f and 1g show the 95% confidence interval Temperature and NODF of these null matrices. Since both the Temperature and NODF of the matrices lie outside the confidence interval, we can say that the nestedness of the matrix is statistically significant.

## Dynamic Null Model

To show that nestedness of the network connecting countries to the products is stable over time we introduce a dynamic null model. This dynamic null model preserves the exact density of the network and also the number of links that appeared and disappeared each year in each country and each product. First, we calculate the number of links that appeared and disappeared for each year. Then, starting with data for the year 1985, we introduced the same number of appearances and disappearances that were observed in the transition between 1985 and 1986 with a location in the matrix determined by the Bascompte et al. null model explained above. The result is a matrix for year 1986 that has the same density of the real data. We continue this procedure to the last year of our data. The procedure was repeated 100 times, and for each matrix we calculated the Temperature and NODF. Figures 1f and 1g of the main text show the 95% confidence interval of the distribution of Temperature and NODF of these dynamic null matrices. The figures show that the dynamic null model does not keep the same level of order of the real data and disorders rather quickly. Hence, the order of the real-data remains highly nested despite large changes in the links of the network.

## DIVERSITY AND UBIQUITY LINES, AND DISTANCE OF EVENTS

To gauge the position of appearances and disappearances in the presence absence matrix, we introduce the diversity and ubiquity lines as a line indicating where presences would be expected to end if the matrix were to be perfectly nested.

In an adjacency matrix sorted by the sum of its rows and columns, the diversity line is a line that goes through the column that is equal to the number of presences in that row. In the case of locations (countries or municipalities) this is equal to their diversity. For each column, the ubiquity line is one that goes through the row equal to its number of presences. In the case of an industry, this represents its ubiquity, or the number of locations where it is present. Figure **SM5** illustrates the diversity and ubiquity lines, and the distance of an event to them.

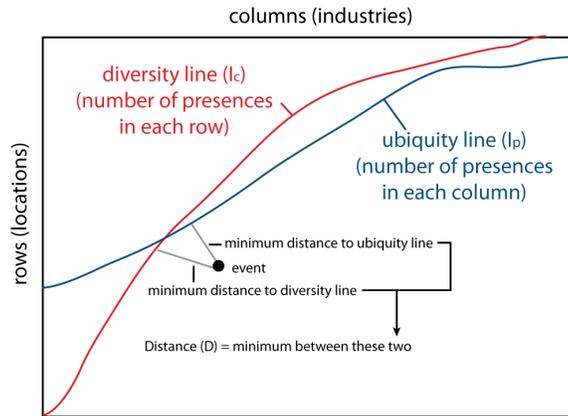

Figure **SM5**: Diversity and Ubiquity lines and the distance of an event to them.

**ROBUSTNESS CHECKS FOR NESTEDNESS**

In this section we show the robustness of some of the main stylized facts of the paper to a different definition of presences and absences. Here, we indicate presences and absences using Balassa's (1986) definition of Revealed Comparative Advantage (RCA). Moreover, we use data for all years (1962-2009).

Balassa's (1986) RCA compares the share of a country's exports that a product represents with the share of world trade represented by that same product. If that product represents a share of that country's export that is larger than its share of world trade, then we say that the country has RCA in that product. We define a presence as having RCA≥1 in a product for at least five consecutive years. Figure SM6 a shows the increase in the number of links in the presence-absence matrix of the RCA network between 1966 and 2009. Figure SM6 b shows the RCA country-product network and their respective diversity and ubiquity line for the year 2000. This matrix is characterized by a temperature of 12±2 and a NODF of 21±8. Figure SM6 c shows its respective Bascompte et al. (2003) null model. In this case, temperature is 12±2 and NODF is 21±8.

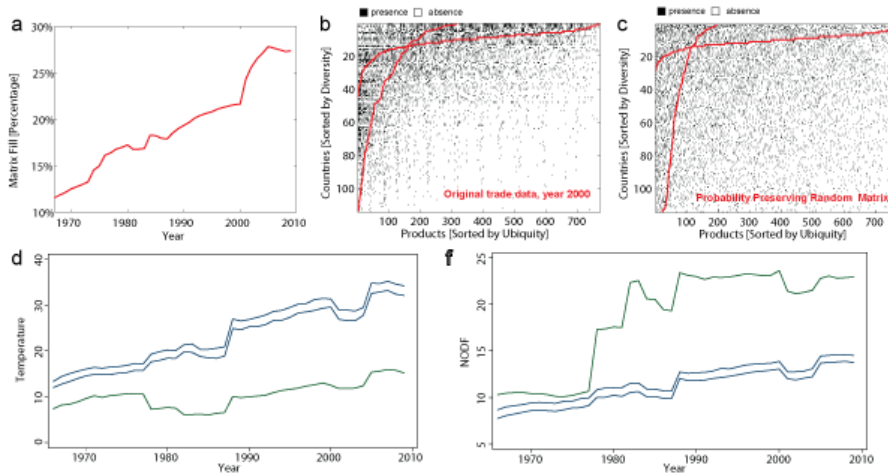

**Figure SM6** The nestedness of international economies using RCA. **a** Evolution of the density of the country-product network between 1985 and 2009. **b** Country-product network for the year 2000. **c** Bascompte el al. null model for the matrix shown in **b**. **d** Evolution of the temperature of the country-product network between 1966 and 2009 (green), its corresponding Bascompte et al. null model (blue, upper and lower lines indicate 95% conf. intervals). **f** Same as **d** but using NODF.

Figure SM7 reproduces Figure 2 of the paper's main text using Balassa's (1986) definition of RCA. These figures illustrate the robustness of the analysis to the difference in definition. It is worth noting that using Balassa's (1986) definition of RCA, instead of the exports per capita definition used in the main text, provides slightly weaker, albeit statistically significant, predictions.

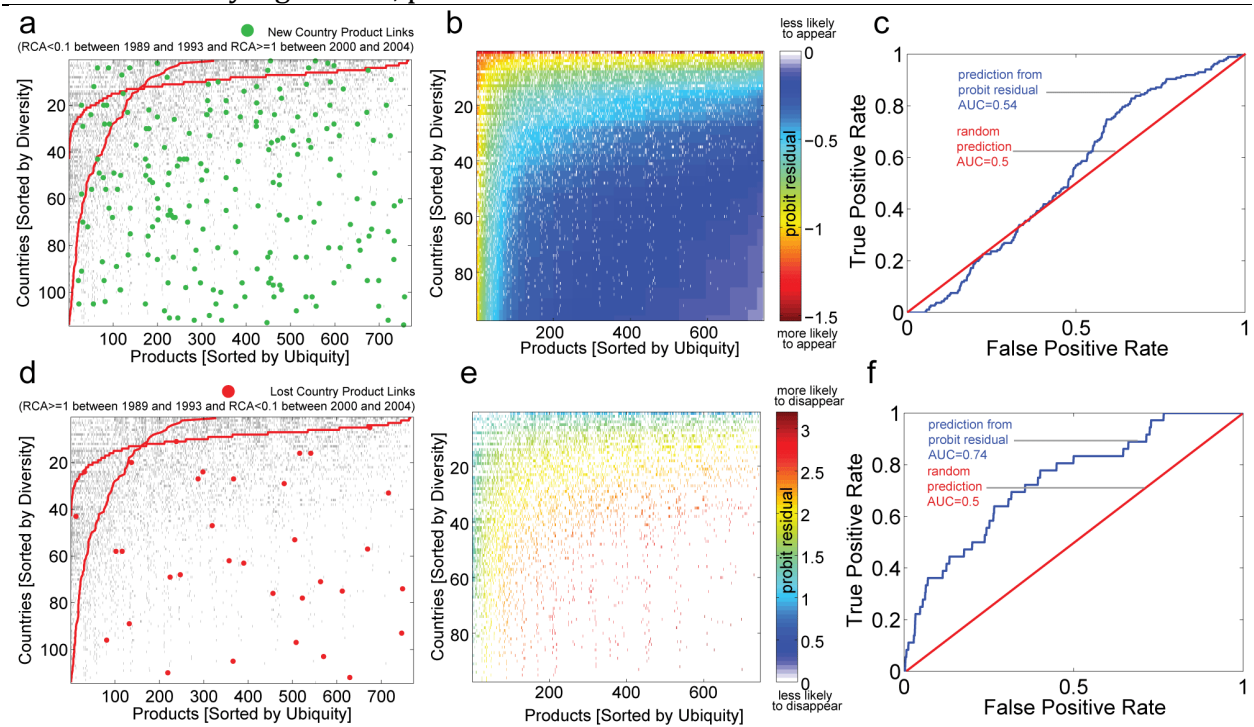

**Figure SM7** Nestedness using RCA. **a** The country-product network for the year 1993 is shown in grey. Green dots show the location of industries that were observed to appear between 1993 and 2000. **d** Same as **a**, but with the industries that disappeared in that period shown in Orange. **b and e** Deviance residuals of the regression presented in (1) of the main text applied to the presences-absences shown in **a-d**. **c and f** ROC curves summarizing the ability of the deviance residuals shown in **b-e**, to predict the appearances and disappearances highlighted in **a and d**.

Finally, Figure SM8 reproduces figure 3 of the main text using Balassa's (1986) RCA to indicate presences and using data for all years. Here, we see that results hold except when the years 1974-1977 are used as predictors. This is because of a large discontinuity in the data classification introduced between 1973 and 1974. This is documented in the second supplementary material of the paper, which shows the fraction of countries that had >0 exports in each product category for all years for the 1006 product categories in the SITC4 rev2 classification.

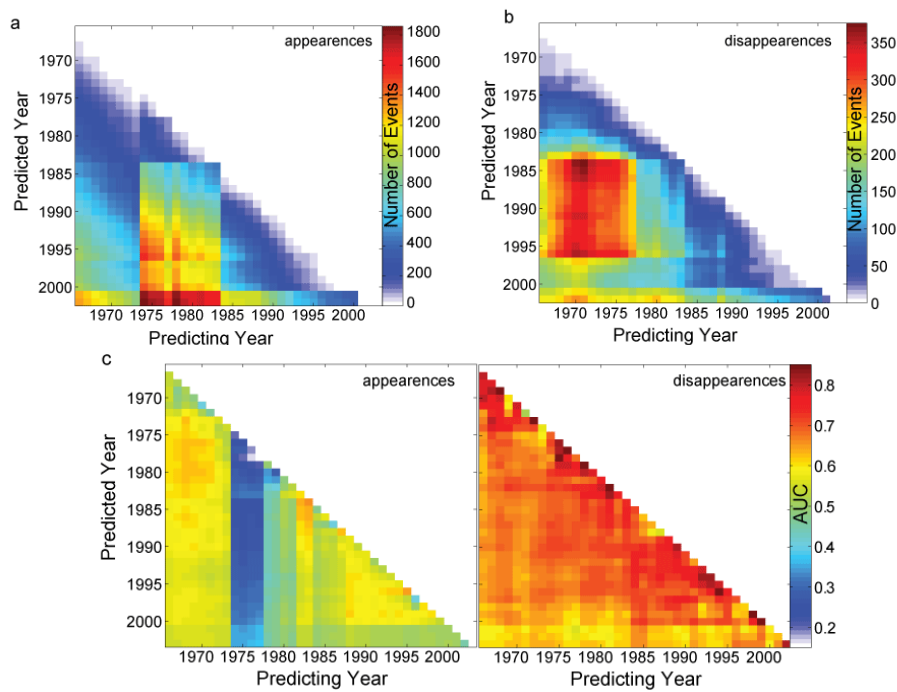

**Figure SM8** Predicting appearances and disappearances using nestedness. **a** Number of appearances for every pair of years in the country-product network. **b** Number of disappearances for every pair of years for the country-product network. **c** Accuracy of the predictions for each pair of years measured using the Area Under the ROC Curve (AUC).

**References:**

Almeida-Neto, M., Guimaraes, P., Guimaraes, P. R., Loyola, R. D. & Ulrich, W. A consistent metric for nestedness analysis in ecological systems: reconciling concept and measurement. *Oikos* **117**, 1227-1239, doi:10.1111/j.2008.0030-1299.16644.x (2008).

Atmar, W. & Patterson, B. D. The Measure of Order and Disorder in the Distribution of Species in Fragmented Habitat. *Oecologia* **96**, 373-382, doi:10.1007/bf00317508 (1993).
B. Balassa, The Review of Economics and Statistics, 68, 315 (1986).

Bascompte, J., Jordano, P., Melian, C. J. & Olesen, J. M. The nested assembly of plant-animal mutualistic networks. *Proceedings of the National Academy of Sciences of the United States of America* **100**, 9383-9387, doi:10.1073/pnas.1633576100 (2003).

Feenstra, R. R. Lipsey, H. Deng, A. Ma and H. Mo. "World Trade Flows: 1962-2000" NBER working paper 11040. National Bureau of Economic Research, Cambridge MA (2005).
Ulrich, W., Almeida, M. & Gotelli, N. J. A consumer's guide to nestedness analysis. *Oikos* **118**, 3-17, doi:10.1111/j.1600-0706.2008.17053.x (2009).